\documentclass[twocolumn,showpacs,preprintnumbers,amsmath,amssymb,superscriptaddress]{revtex4}
\usepackage{graphicx}
\usepackage{bm}

\begin{document}
\title{Two-soliton collisions in a near-integrable lattice system}
\author{ S.~V.~Dmitriev}
\affiliation{Institute of Industrial Science, the University of Tokyo, 
4-6-1 Komaba, Meguro-ku, Tokyo 153-8505, Japan}

\author{ P.~G.~Kevrekidis}
\affiliation{Department of Mathematics and Statistics, University of Massachusetts, 
Amherst, Massachusetts 01003-4515}

\author{ B.~A.~Malomed}
\affiliation{Department of Interdisciplinary Studies, 
Faculty of Engineering, Tel Aviv University, Tel Aviv 69978, Israel}

\author{ D.~J.~Frantzeskakis}
\affiliation{Department of Physics, University of Athens Panepistimiopolis, 
Zografos, Athens 15784, Greece}

\begin{abstract}
We examine collisions between identical solitons in a weakly perturbed
Ablowitz-Ladik (AL) model, augmented by either onsite cubic nonlinearity
(which corresponds to the Salerno model, and may be realized as an array of
strongly overlapping nonlinear optical waveguides), or a quintic
perturbation, or both. Complex dependences of the outcomes of the collisions
on the initial phase difference between the solitons and location of
the collision point are observed. Large changes of amplitudes and velocities
of the colliding solitons are generated by weak perturbations, showing that
the elasticity of soliton collisions in the AL model is fragile (for instance, the Salerno's
perturbation with the relative strength of $0.08$ can give rise to a change of
the solitons' amplitudes by a factor exceeding $2$). Exact and approximate
conservation laws in the perturbed system are examined, with a conclusion
that the small perturbations very weakly affect the norm and energy
conservation, but completely destroy the conservation of the lattice
momentum, which is explained by the absence of the translational
symmetry in generic nonintegrable lattice models. Data collected for a very
large number of collisions correlate with this conclusion. Asymmetry of
the collisions (which is explained by the dependence on the
location of the central point of the collision relative to the lattice, and
on the phase difference between the solitons) is investigated too, showing
that the nonintegrability-induced effects grow almost linearly with
the perturbation strength. Different perturbations (cubic and quintic ones)
produce virtually identical collision-induced effects, which makes it
possible to compensate them, thus finding a special perturbed system with
almost elastic soliton collisions.
\end{abstract}


\maketitle

\vspace{-15mm}

\section{Introduction}

Soliton collisions is one of central topics of nonlinear-wave dynamics. In
integrable systems, solitons are well known to emerge unscathed from
collisions \cite{AS}. However, even small nonintegrable perturbations may
render the phenomenology much richer, causing various inelastic effects,
such as trapping and formation of bound states, multiple-bounce interactions
(where solitons separate after multiple collisions) \cite{campbell},
fractality in the outcome of the collision \cite{anninos,Yang}, and others.
Such complex features are usually attributed to the excitation of soliton
internal modes \cite{campbell,Yang,pgk}, but more recently it was realized
that they also occur due to the possibility for radiationless energy
exchange (even in the absence of internal modes) between the colliding
solitons, should the conservation laws allow it \cite{inelasticPRE,Chaos}.
The latter mechanism was both confirmed by direct simulations of the
corresponding nonintegrable models, and might be expected to ensue from the
principle stating that any outcome compatible with the conservation laws may
take place under appropriate initial conditions. 

Such effects suggest that the integrability is essentially tantamount to the
strictly elastic character of the collisions \cite{AS}, and warrant the
importance of further studies of strongly inelastic collision effects
produced by small conservative perturbations added to basic integrable
models. This general issue is of interest not only in its own right, but
also for applications to nonlinear optical waveguides, as strong changes in
the character of the interaction induced by a small perturbation may be
naturally used in the context of switching, see e.g., \cite{meier} and
references therein.

The objective of the present work is to consider such effects in collisions
of dynamical (rather than topological) solitons in a \emph{discrete} 
near-integrable system. As a matter of fact, the only integrable system which can
be used in this case as the zeroth-order approximation is the Ablowitz-Ladik
(AL) lattice \cite{AL}. Its well-known nonintegrable extension is the
Salerno model (SM) \cite{Salerno}, which is produced by adding the
integrability-breaking perturbation in the form of the onsite nonlinearity
to the integrable AL system with the inter-site cubic nonlinearity. In order
to test if the results that will be obtained below are generic, we will also
consider an essentially different type of an integrability-breaking
conservative perturbation, viz., the quintic onsite nonlinearity [its
principal difference from the cubic counterpart is that it breaks the
integrability of both the AL lattice and of its continuum limit, i.e., the
nonlinear Schr\"{o}dinger (NLS) equation].

Thus, we introduce a general dynamical model based on the following equation:
\begin{eqnarray}
&&i\dot{\psi}_{n}+\left( 2h^{2}\right) ^{-1}(\psi _{n-1}-2\psi _{n}+\psi
_{n+1})+\delta |\psi _{n}|^{2}\psi _{n}  \nonumber \\
+(1/2)\left( 1-\delta \right) |\psi _{n}|^{2}(\psi _{n+1}+\psi _{n-1})
&=&\varepsilon |\psi _{n}|^{4}\psi _{n},  \label{sal1}
\end{eqnarray}
Here $\psi _{n}$ is the complex dynamical variable at the $n$-th site of the
lattice, the overdot stands for the time derivative, $h$ is the lattice
spacing, $\varepsilon $ is a real constant controlling the quintic
perturbation, and $\delta $ is a real parameter that accounts for the
crossover between the AL ($\delta =0$, $\varepsilon =0$) and discrete-NLS 
($\delta =1$, $\varepsilon =0$) \cite{KRB} limits. Equation (\ref{sal1})
conserves two dynamical invariants, namely, the norm, 
\begin{equation}
\mathcal{N}=\frac{1}{1-\delta }\sum_{n}\ln \left[ 1+h^{2}(1-\delta )|\psi
_{n}|^{2}\right] ,  \label{no}
\end{equation}
and energy (Hamiltonian), 
\begin{eqnarray}
\mathcal{H} &=&-\sum_{n}\left\{ \frac{h^{2}(1-\delta )+\varepsilon }{
h^{2}(1-\delta )^{3}}\ln \left[ 1+h^{2}(1-\delta )|\psi _{n}|^{2}\right]
\right.   \nonumber \\
&&-\frac{h^{2}(1-\delta )+\varepsilon }{(1-\delta )^{2}}|\psi _{n}|^{2}+
\frac{h^{2}}{2}|\psi _{n}-\psi _{n-1}|^{2}  \nonumber \\
&&+\left. \frac{\varepsilon h^{2}}{2(1-\delta )}|\psi _{n}|^{4}\right\} .
\label{H}
\end{eqnarray}

While the discrete NLS equation, corresponding to $\delta =1$ and 
$\varepsilon =0$ in Eq. (\ref{sal1}), has numerous physical realizations, the
most important one being arrays of nonlinear optical waveguides 
\cite{Yaron}, the AL model does not directly apply to many 
physical systems, because of
the specific character of the nonlinear terms in it. However, a realization
of the SM may be an array of \emph{strongly overlapping} nonlinear optical
waveguides, especially the one following a zigzag pattern (similar to an
array introduced in Ref. \cite{zigzag}). Indeed, the overlapping between
adjacent cores will give rise, through the Kerr effect, to a nonlinear
correction in the linear coupling between the cores, in the form of the
terms $\sim \left( 1-\delta \right) $ in Eq. (\ref{sal1}). It should be
noted that, in this case, extra perturbation terms are expected too, such as 
$\left( |\psi _{n-1}|^{2}+|\psi _{n+1}|^{2}\right) \psi _{n}$ (cross-phase
modulation). However, the results presented below clearly demonstrate that
strong effects generated by small conservative perturbations are essentially
the same for different perturbations, therefore we expect that taking into
regard all the possible perturbation terms corresponding to the optical
waveguides with strong overlap between the cores will not alter the results
significantly.

In some specific cases, soliton collisions in the SM have already been
examined. In particular, a collision between a soliton and a reflecting
wall, which is equivalent to a strictly symmetric collision between the
soliton and its mirror image, were studied numerically in Ref. \cite{Cai}.
One of our aims is to explore sensitivity of collisions to small asymmetries
in initial phases and positions of the solitons in the actual two-soliton
collision. Very recently, collisions in the (strongly nonintegrable)
discrete NLS equation were examined \cite{ioanna}, and symmetry-breaking
effects were found, along with sensitivity of the outcome to the location of
the collision point. Here, we present results of collisions and their
dependence on parameters in the model (\ref{sal1}) with small $\delta $ and 
$\varepsilon $, i.e., close to the integrable AL limit. Together with the
already available findings for the strongly nonintegrable case 
\cite{ioanna}, they provide for a sufficiently comprehensive description of the
collisional dynamics of nontopological solitons in fundamental lattice
systems.

The AL model [Eq. (\ref{sal1}) with $\delta =0$ and $\varepsilon =0$] has
exact soliton solutions of the form 
\begin{equation}
\psi _{n}(t)=\frac{1}{h}\frac{\sinh \mu }{\cosh [\mu (n-x(t))]}\exp \left\{
ik\left[ n-x(t)\right] +i\alpha (t)\right\} \,,  \label{sal2}
\end{equation}
where the instantaneous coordinate and phase of the soliton are  
\begin{eqnarray}
x(t) &=&x_{0}+\frac{t}{h^{2}}\left( \sin k\right) \frac{\sinh \mu }{\mu }\,,
\nonumber \\
\alpha (t) &=&\alpha _{0}+\frac{t}{h^{2}}\left[ \left( \cos k\right) \cosh
\mu +k\left( \sin k\right) \frac{\sinh \mu }{\mu }-1\right] \,.
\label{xalpha}
\end{eqnarray}
$x_{0}$ and $\alpha _{0}$ are their initial values, while $\mu $ and $k$
define the soliton's amplitude $A$ and velocity $V$, 
\begin{equation}
A=h^{-2}\sinh ^{2}\mu \,,\,\,\,\,\,\,V=\left( \mu h\right) ^{-1}\left( \sinh
\mu \right) \sin k\,.  \label{sal3}
\end{equation}

The infinitely long AL system has an infinite series of dynamical
invariants, the lowest ones being the norm, lattice momentum, and energy, 
\begin{equation}
N=\sum_{n}\ln (1+h^{2}|\psi _{n}|^{2}),  \label{sal4}
\end{equation}
\begin{equation}
P=ih^{2}\sum_{n}(\psi _{n}\psi _{n+1}^{\ast }-\psi _{n}^{\ast }\psi _{n+1}),
\label{sal5}
\end{equation}
\begin{equation}
Q=\frac{1}{2}h^{2}\sum_{n}(\psi _{n}\psi _{n+1}^{\ast }+\psi _{n}^{\ast
}\psi _{n+1}).  \label{sal6}
\end{equation}
Note that the norm of the general nonintegrable model (\ref{sal1}), given by
the expression (\ref{no}), goes over into the norm (\ref{sal4}) in the limit 
$\delta =\varepsilon =0$, and the Hamiltonian (\ref{H}) of the nonintegrable
model becomes, in the same limit, a linear combination of the norm (\ref
{sal4}) and energy (\ref{sal6}) of the AL integrable system: $H(\delta
=\varepsilon =0)\equiv -\left( N+Q\right) $. It will be seen below that, as
a matter of fact, the difference between the exact norm and Hamiltonian of
the full perturbed model with small $\delta $ and $\varepsilon $ and those
of the AL model is negligible. However, all the other dynamical invariants
of the AL model, \emph{including} the lattice momentum (\ref{sal5}), have no
counterparts in the nonintegrable case. This is explained by the fact that
each elementary dynamical invariant is generated by a certain continuum
invariance of the underlying equation. In particular, the norm and energy
conservation are accounted for by the invariance against phase and time
shifts, respectively, that remain valid in the nonintegrable system, while a
hidden dynamical symmetry of the AL model which is responsible for the
conservation of the lattice momentum is destroyed by the small
perturbations. The momentum remains a dynamical invariant in continuum
nonintegrable models (e.g., the NLS equation with the quintic term), but in
the discrete setting it is conserved solely in the integrable case --
obviously, a generic lattice system is not invariant against arbitrary
spatial translations. Further consideration of this issue can be found in a
recent work \cite{Panos}.

An issue related to the lack of conserved momentum is the (non)existence of
exact traveling soliton solutions in nonintegrable lattice models.
While the problem still waits for its full solution, several important
theoretical results have been obtained. The inclusion of an appropriate traveling
ansatz in the discrete equation gives rise to a differential-delay equation 
whose steady states are the traveling wave solutions of the original 
differential-difference equation (see e.g., Refs. \cite{Iooss,Friesecke},
as well as an earlier work by Feddersen, Ref. \cite{Feddersen}).
It is also pertinent to
mention that moving solitons clearly persist in simulations of perturbed
systems, without any conspicuous loss, for long times, which is sufficient
to study their collisions in numerical experiments without ambiguity (see,
e.g., Refs. \ \cite{Cai} and \cite{OFZ}).

If the given system differs from the AL model by small terms, a natural
question is how strong actual destruction of the former dynamical
invariants, and especially of the momentum, which has a straightforward
physical interpretation (and remains a virtually conserved quantity for free
moving solitons, as explained above) will be in collisions between solitons.
One of main objectives of the present work is to address this issue. We will
conclude that the momentum conservation is \emph{strongly} violated by the
collisions, even if the perturbation parameters are quite small.

For two broadly separated AL solitons with parameters $\mu _{j}$ and 
$k_{j}$, the expressions (\ref{sal4}) -- (\ref{sal6}) take values 
\begin{equation}
N_{\mathrm{sol}}=2\sum\limits_{j=1}^{2}\mu _{j},\,\,\,Q_{\mathrm{sol}
}=2\sum\limits_{j=1}^{2}\left( \sinh \mu _{j}\right) \cos k_{j}.\,\,
\label{NQ}
\end{equation}
\begin{equation}
\,P_{\mathrm{sol}}=2\sum\limits_{j=1}^{2}\left( \sinh \mu _{j}\right) \sin
k_{j},  \label{P}
\end{equation}
The fact that only two exact and, plausibly, one approximate conserved
quantities (the latter one is the momentum) constrain possible outcomes of
the soliton-soliton collisions, which are characterized by two amplitudes,
two velocities, and, in addition, may depend on the initial relative phase 
$\Delta \alpha _{0}=\alpha _{02}-\alpha _{01}$ and positions $\left(
x_{0}\right) _{2}$ and $\left( x_{0}\right) _{1}$ of the solitons, suggests
that the above-mentioned radiationless energy exchange between the two
solitons is quite feasible. Furthermore, for slow solitons (small $k_{j}$)
the conservation of $Q$ becomes an amplitude constraint [see Eqs. (\label{sal3})],
and if amplitudes are small too ($\mu _{j}\rightarrow 0$), $Q$ reduces
to $N$, see Eqs. (\ref {NQ}), so that there actually remains the single 
constraint in this limit case.

\section{Numerical results}

\subsection{Setting up the problem}

To perform simulations of the collisions, we notice that $h$ can actually be
scaled out from Eq. (\ref{sal1}), leaving $\delta $ and $\varepsilon /h^{2}$
as independent control parameters, therefore in what follows, we fix 
$h=0.8$. The parameters are varied in ranges corresponding to the weak quintic
perturbation, $\varepsilon \in \lbrack -0.01,0.01]$, and moderately weak
Salerno's perturbation, $\delta \in \lbrack -0.08,0.08]$. Equations (\ref
{sal1}) were integrated by means of an implicit Crank-Nicholson scheme with
the accuracy of $O(\left( \Delta t\right) ^{2})$ and with reflecting
boundary conditions. The initial condition was taken as a superposition of
two far separated solitons (\ref{sal2}) that would be exact solutions of the
AL model, and the numerical integration was run until outgoing solitons were
separated well enough. Their amplitudes and velocities after the collision, 
$\tilde{A}_{1}$, $\tilde{A}_{2}$, $\tilde{V}_{1}$, and $\tilde{V}_{2}$, were
measured, and then the corresponding parameters $\tilde{\mu}_{1}$, $\tilde{
\mu}_{2}$, $\tilde{k}_{1}$, and $\tilde{k}_{2}$ were found inverting Eqs. 
(\ref{sal3}). We also checked to what extent the dynamical invariants given
by the unperturbed expressions (\ref{NQ}) and (\ref{P}), as well as by the
exact ones (\ref{no}) and (\ref{H}), were conserved.

To present the results, we will focus on the symmetric collisions: $\mu
_{1}=\mu _{2}=\mu $, $k_{1}=-k_{2}=k$. While simulations were run for
various values of the amplitudes and velocities, we display results for a
case that turned out to be a typical one, adequately representing many
others, for $\mu =0.75$, $k=0.1$. This implies $A_{1}=A_{2}=A=1.057$, 
$V_{1}=-V_{2}=V=0.137$. The initial phase difference $\Delta \alpha _{0}$ was
controlled by setting $\alpha _{01}=0$ and choosing $\alpha _{02}$ from the
interval $(-\pi ,\pi )$. The initial positions of the solitons were taken as 
$(x_{0})_{1}=-x_{0}+x_{c}$ and $(x_{0})_{2}=x_{0}+x_{c}$, with $x_{0}=12$;
this provides for the large initial separation $2x_{0}=24$ between them,
while $x_{c}$ was chosen from the interval $[0,1)$ to control the location
of the collision point.

The presentation of results is structured as follows: we first examine the
effect of variation of the initial phase difference $\Delta \alpha _{0}$ and
collision point $x_{c}$ on the outcome of the collision. Then, we analyze
how the approximate conservation of the expressions (\ref{NQ}) and (\ref{P})
correlates with the results. Finally, we examine the effect of combining the
perturbation parameters $\varepsilon $ and $\delta $, in order to
demonstrate that the two perturbations may almost exactly cancel each other,
thus making the collisions virtually elastic. We stress that results
obtained at other values of parameters are completely tantamount to those
displayed below, provided that $\varepsilon $ and $\delta $ remain small.

\subsection{Sensitivity to the phase and position of the collision}

\begin{figure}[th]
\includegraphics[scale=1.3]{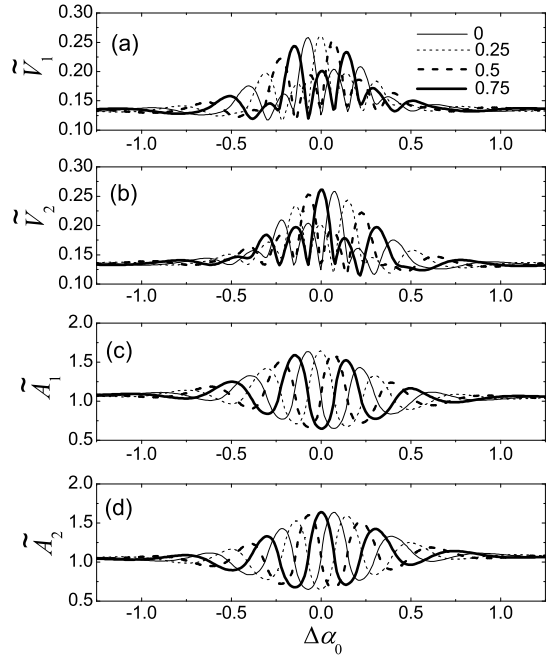}
\caption{Velocities and amplitudes of the solitons after the collision vs.
the phase difference $\Delta\protect\alpha_0$, in the case 
$\delta=0.04$, $\protect\varepsilon=0$ (the Salerno model without quintic terms).
This and all the other cases are shown for $\protect\mu=0.75$ and $k=0.1$,
see Eq. (\ref{sal3}). Four different curves correspond to different
positions of the collision point, $x_c=0$, $0.25$, $0.5$, and $0.75$. The
collisions are strongly inelastic in the vicinity of 
$\Delta\alpha_0=0$.}
\label{sfig1}
\end{figure}

In Fig. \ref{sfig1}, values of the soliton parameters after the collision
are presented as functions of the initial phase difference $\Delta \alpha
_{0}$ for the case of the SM perturbation. In the part of the interval 
$\left( -\pi ,\pi \right) $ which is not included, the collision is almost
completely elastic. It is obvious that, in the interval $\left| \Delta
\alpha _{0}\right| \,_{\sim }^{<}\,\,0.5$, the small perturbation is, in
fact, a singular one, resulting in very strong effects (in other words, the
elasticity of soliton collisions in the AL lattice is a very fragile feature). Note that, in
the case of a weakly perturbed \emph{continuum} NLS equation considered in
Ref. \cite{inelasticPRE}, noteworthy inelastic effects in the collision of
two solitons also took place at relatively small values of $\left| \Delta
\alpha _{0}\right| $.

\subsection{Dynamical invariants}

\begin{figure}[th]
\includegraphics[scale=1.25]{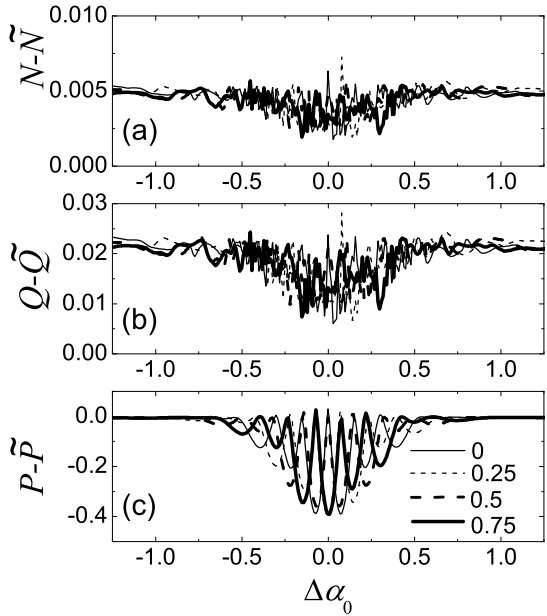}
\caption{The collision-induced changes of the net norm, energy, and momentum
for the two solitons, defined as per Eqs. (\ref{NQ}) and (\ref{P}), vs. 
$\Delta\protect\alpha_0$ for different values of $x_c$ in the Salerno model
(recall that the tilde refers to the post-collision values of the
corresponding quantities). The quantities displayed in this figure are
obtained by adding up their values for the two solitons
(rather than by direct calculation for the whole system). If the norm
and energy are defined by the exact expressions (\ref{no}) and 
(\ref{H}), rather than the approximate ones (\ref{NQ}), they are completely
conserved.}
\label{sfig2}
\end{figure}

In the present case, the initial values of the expressions (\ref{NQ}) and 
(\ref{P}) are $N=3$, $Q=3.273$, and $P=0$ (the absolute values of the initial
momenta are $|P_{1}|=|P_{2}|=0.1642$). Using the exact expressions 
(\ref{no}) and (\ref{H}), we have checked that the net values of the norm and energy
for the initial solitons and those observed after the collision are equal,
in the case of the ordinary numerical accuracy employed, up to $10^{-4}$ (in
relative units); running simulations with higher accuracy (smaller $\Delta t$), it was possible
to check the norm and energy conservation with the accuracy of up to $10^{-6}
$. Norm and energy loss due to radiation loss remained completely negligible
in all the cases considered. As concerns the difference between the
unperturbed expressions used in Eqs. (\ref{NQ}) and the exact ones 
(\ref{no}) and (\ref{H}), Fig. \ref{sfig2} demonstrates that the largest relative
difference between them, which reflects a direct effect of the small
perturbations, is $\Delta N/N\sim 10^{-3}$ for the norm, and $\Delta Q/Q\sim
3\times 10^{-3}$ for the energy. However, the bottom panel in Fig. \ref
{sfig2} shows that the momentum is \emph{not} conserved in any
approximation, in accordance with the fact that the perturbed system has no
translational symmetry.

The conservation of $N$ suggests that a simple relation between the soliton
amplitudes after the collision may be expected: according to Eqs. 
(\ref{NQ}), $\mu _{1}+\mu _{2}$ must keep the original value with the accuracy 
$\sim 10^{-3}$. On the contrary, the momentum nonconservation promises a much
worse accuracy in the prediction of a relation between the velocities. The
conservation of $Q$ does not provide for an essential additional information
for small values of $k$ (see above), while large $k$ implies the collision
between fast solitons, when nontrivial effects will be very weak. 

\begin{figure}[th]
\includegraphics[scale=1.5]{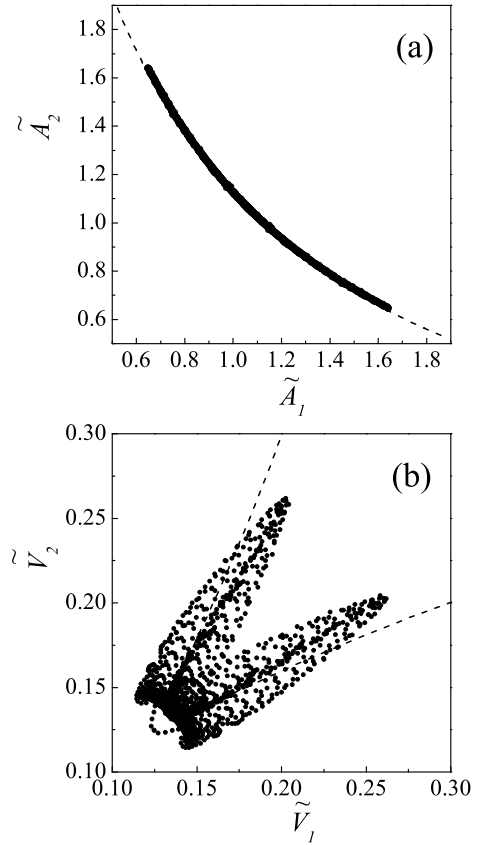}
\caption{Relations between the solitons' amplitudes (a) and velocities (b)
after the collision in the Salerno model. The dashed lines show the
relations predicted by the norm, energy, and momentum conservation
for integrable AL chain Eqs. (\ref{NQ}) and (\ref{P}). Dots are
numerical results for 2500 collisions, with values of the initial phase
difference $\Delta\alpha_0$ taken from the interval $[-1.25,1.25]$
with a step of $0.01$, and the collision-point's coordinate $x_c$ taken from 
$[0,1)$ with a step of $0.1$. Note that the norm was taken in the
approximate form of Eq. (\ref{NQ}), which pertains to the unperturbed
Ablowitz-Ladik lattice. The relative nonconservation of this norm after the
collision is $\sim 10^{-3}$ (see the text), while the exact norm of the
perturbed model, as given by Eq. (\ref{no}), is conserved exactly, within
the numerical accuracy.}
\label{sfig3}
\end{figure}

The comparison between the actual results of the collision (dots) and
predictions based on the approximate conservation laws for the values $N$
and $P$ of the two solitons in the form of Eqs. (\ref{NQ}) (dashed lines) is
displayed in Fig. \ref{sfig3}. As is seen, the amplitude relation indeed
follows from the norm conservation in a very accurate form, while the
conservation of the momentum may be traced in a very crude form only.

\begin{figure}[th]
\includegraphics[scale=0.9]{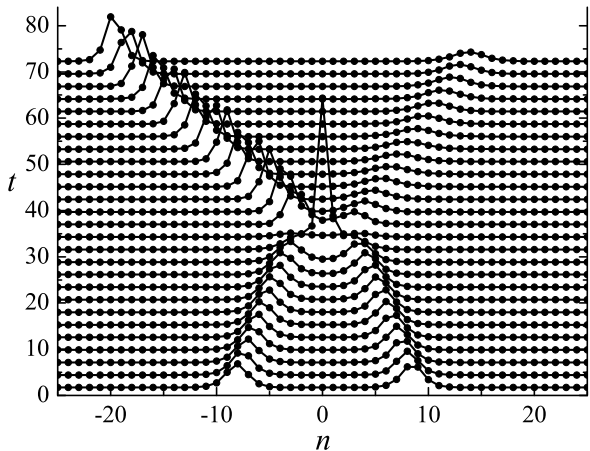}
\caption{An example of a strongly inelastic collision between solitons in
the AL system with a small on-site quintic perturbation, $\protect\delta =0$
and $\varepsilon =-0.01$. In this case, $\Delta \alpha
_{0}=0 $ and $x_{c}=0.2$.}
\label{overall}
\end{figure}

Another salient feature of Fig. \ref{sfig3} is strong deviation of the dots
from the diagonal point $(1.054,1.054)$ corresponding to the values of the
parameters before the collision. Such a feature was impossible in the case
of the collision of a soliton with its mirror image in the SM, examined in\
Ref. \cite{Cai}. A typical example of an inelastic collision (inducing this
effect) is shown in Fig. \ref{overall}. The major cause of the effect is the
location of the collision central point $x_{c}$ relative to the underlying
lattice. 

Besides that, the phase difference between the colliding solitons may
produce a similar symmetry-breaking effect. Indeed, if the AL solitons
described by Eqs. (\ref{sal2}) and (\ref{xalpha})
moving to the right and to the left [with $k>0$ and $k<0$, accordingly],
are given phase shifts $+\Delta \phi $ and $-\Delta \phi$,
this is equivalent to the shift of the coordinate $x$, but \emph{solely}
in the expressions for the solitons' phases, by $\Delta x=\phi _{0}/k$,
which has \emph{equal} signs for both solitons. This means the phase pattern
of the two-soliton configuration gets shifted by $\Delta x$ relative to the
shapes of the colliding solitons, which is an obvious cause for the symmetry
breaking. The fact that the above-mentioned position and phase factors do
not affect the collision symmetry in the integrable AL model is another specific
manifestation of its integrability.

\subsection{The role of the perturbation strength}

\begin{figure}[th]
\includegraphics[scale=0.8]{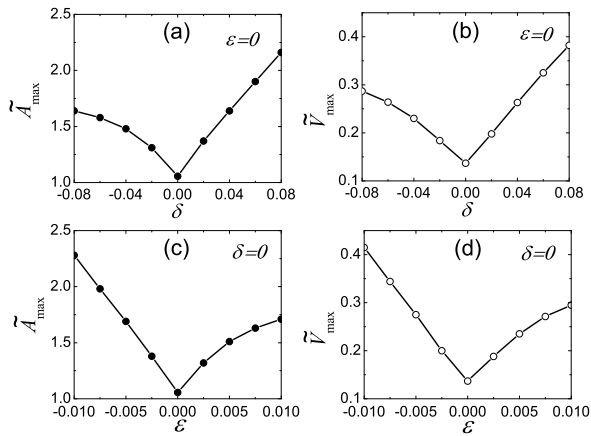}
\caption{The maximum possible amplitude $\tilde A_{\mathrm{max}}$ (filled
circles) and velocity $\tilde V_{\mathrm{max}}$ (empty circles) of the
soliton after the collision vs. the perturbation strengths $\protect\delta$
and $\protect\varepsilon$: (a,b) $\protect\varepsilon=0$ (the Salerno
model); (c,d) $\protect\delta=0$ (the quintic model).}
\label{sfig4}
\end{figure}

In Fig. \ref{sfig3} one can observe that, for $\delta =0.04$ and 
$\varepsilon =0$, the maximum possible soliton amplitude after the collision
is $\tilde{A}_{\mathrm{max}}=1.64$, and the maximum possible post-collision
velocity is $\tilde{V}_{\mathrm{max}}=0.263$. These values [and, in
particular, their deviation from initial ones, $(A,V)=(1.054,0.137)$] may be
regarded as a measure of the departure of the perturbed model from the
integrability. In Fig. \ref{sfig4}, we use $\tilde{A}_{\mathrm{max}}$ and 
$\tilde{V}_{\mathrm{max}}$ to gauge the deviation from the integrable case
with the increase of the perturbation strength (the cases of both the SM and
quintic perturbations are shown). It is concluded that the weak
perturbations generate quite large inelastic effects, the inelasticity
increasing almost linearly with the perturbation parameter. For instance,
the value $\delta =+0.08$ of the relative perturbation parameter in the SM
model gives rise to collision-induced changes of the soliton's amplitude by
a factor of $\simeq 2$, and of the absolute value of the velocity by a
factor of $\simeq 3$.

Another noteworthy feature is the asymmetry of the plots in Fig. 
\ref{sfig4}. The asymmetry is due to the fact that internal modes in the colliding
solitons can be excited only when $\delta >0$ (for $\varepsilon =0$) or when 
$\varepsilon <0$ (for $\delta =0$) \cite{pelinov}. Hence, in these cases, we
observe a combined effect of the radiationless energy exchange and the
internal-mode excitation, while for $\delta <0$, $\varepsilon =0$ and 
$\delta =0$, $\varepsilon >0$, only the former occurs. Naturally, the net
nonintegrability-induced effects are stronger in the cases where the
internal mode can be excited.

\subsection{Compensation of perturbation effects}

\begin{figure}[th]
\includegraphics[scale=1.1]{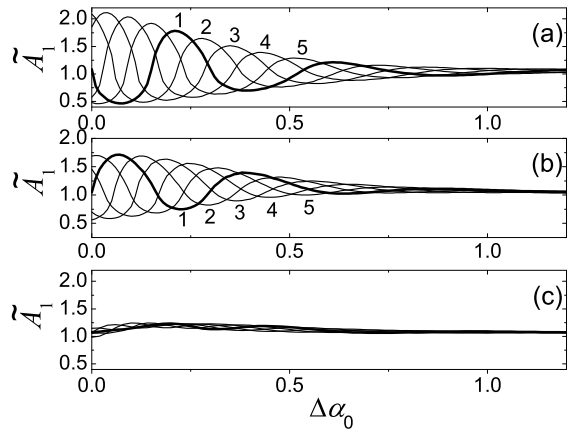}
\caption{The amplitude of the first soliton after the collision, 
$\tilde A_1$, vs. $\Delta\alpha_0$ for $x_c=0$, $0.2$, $0.4$, $0.6$, and $0.8$,
curves 1 to 5, respectively. The perturbations have (a) $\delta=0.08$,
$\varepsilon=0$; (b) $\protect\delta=0$, $\protect\varepsilon=0.01$;
(c) $\protect\delta=0.08$, $\protect\varepsilon=0.01$.}
\label{sfig5}
\end{figure}

In Ref. \cite{inelasticPRE}, it was found that, in continuum models,
inelastic effects in soliton collisions can be strongly suppressed if
contributions from different perturbations cancel each other. We have
observed a similar feature in the present model. In particular, in Fig. \ref
{sfig5} we show the post-collision amplitude $\tilde{A}_{1}$ versus $\Delta
\alpha _{0}$ for three different perturbations. It is clear that the
cancellation takes place in the case (c), making the norm and momentum
exchange an order of magnitude smaller than in the other cases. Similar
compensation effects were observed for $\delta=0.08$ and
$\varepsilon=0.01$ in a wide range of soliton parameters, including
non-symmetric collisions (between non-identical solitons). In fact, the
possibility of the mutual compensation between the Salerno and quintic
perturbation is a strong proof to the assertion that different conservative
perturbations produce virtually identical effects, hence essentially the
same results are expected from other perturbations.

\section{Discussion and Conclusions}

In this work, we have quantified properties of collisions between solitons
in the Ablowitz-Ladik (AL) model with weak Hamiltonian perturbations. We
have observed complex dependences of the outcomes of the collisions on the
initial phase difference between the solitons and exact location of the
collision point. Strong inelastic effects, in the form of radiationless
energy and momentum exchange between colliding solitons, are generated by
weak perturbations (for instance, a perturbation with the relative strength 
$\delta =0.08$ gives rise to a change of the solitons' amplitudes by a factor
exceeding $2$). The effects produced by different conservative perturbations
are quite similar, suggesting that the results reported in this paper are
generic. The exact and approximate conservation laws of the perturbed system
were examined, with a conclusion that the small perturbations very weakly
affect the norm and energy conservation, but strongly destroy the
conservation of the lattice momentum, which is explained by the absence
of the translational symmetry in nonintegrable lattice models. Statistical
data collected for a very large number of collisions validate this
conclusion. Symmetry-breaking effects in the collisions (which are simply
explained by the dependence of the result on the location of the central
point of the collision relative to the lattice, and by the phase difference
between the colliding solitons) were highlighted, and their magnitude was
used to gauge the deviation of the perturbed model from integrability. It
was also shown that, properly combining two different perturbations, it is
possible to almost exactly cancel their integrability-destroying effects,
thus constructing a perturbed system in which collisions are practically
elastic.

In this paper, we were dealing with collisions between solitons with
relatively large initial velocities. It would naturally be of interest to
see how the picture presented is modified for smaller collision velocities,
and, in particular, to examine whether a fractal structure, similar to that
observed in Ref. \cite{Chaos}, can be found in the present model. This issue
will be considered elsewhere.

\acknowledgments 

We appreciate a discussion with M. Salerno. PGK gratefully acknowledges
support from NSF-DMS-0204585. BAM appreciates hospitality of the Department
of Mathematics at the University of Massachusetts (Amherst), and financial
support from the European Office of Aerospace Research and Development (US\
Air Force), provided through a Window-on-Science grant.

\end{document}